\documentstyle[12pt]{article}

\newcommand{\be}{\begin{equation}}\newcommand{\ee}{\end{equation}}
\newcommand{\bea}{\begin{eqnarray}}\newcommand{\eea}{\end{eqnarray}}
\newcommand{\nn}{\nonumber}\newcommand{\p}[1]{(\ref{#1})}

\begin{document}
\begin{titlepage}
\begin{flushright}
ENSLAPP-L-623/96  \\
JINR E2-96-394 \\
hep-th/9611033 \\
November 1996
\end{flushright}
\vskip 1.0truecm
\begin{center}
{\large \bf NEW SUPER KdV SYSTEM WITH THE N=4 SCA \\
AS THE HAMILTONIAN 
STRUCTURE}
\end{center}
\vskip 1.0truecm
\centerline{\bf F. Delduc${}^{(a)}$, L. Gallot${}^{(a)}$, 
E. Ivanov${}^{(b)}$}
\vskip 1.0truecm
\centerline{${}^{(a)}$ \it Laboratoire de Physique Th\'eorique ENSLAPP
\footnote[1]{URA 1436 du CNRS associ\'ee
\`a l'Ecole Normale Sup\'erieure de Lyon et \`a
l'Universit\'e de Savoie}, ENS Lyon}
\centerline{\it 46 All\'ee d'Italie, 69364 Lyon, France}
\vskip5mm
\centerline{${}^{(b)}$\it Bogoliubov Laboratory of Theoretical 
Physics, JINR,}
\centerline{\it Dubna, 141 980 Moscow region, Russia}
\vskip 1.0truecm  \nopagebreak

\begin{abstract}
We present a new integrable extension of the $a=-2$, $N=2$ SKdV hierarchy, 
with the "small" $N=4$ superconformal algebra (SCA) as the second 
hamiltonian structure. As distinct 
from the previously known $N=4$ supersymmetric KdV hierarchy associated with 
the same $N=4$ SCA, the new system respects only $N=2$ rigid 
supersymmetry. We give for it both matrix and scalar Lax formulations 
and  consider its various integrable reductions which complete the list 
of known SKdV systems with the $N=2$ SCA as the second hamiltonian structure. 
We construct a generalized Miura transformation which relates our 
system to the $\alpha = -2$, $N=2$ super Boussinesq 
hierarchy and, respectively, the ``small'' $N=4$ SCA to the $N=2$ $W_3$ 
superalgebra. 
\end{abstract}
\end{titlepage}

\section{Introduction}
In the recent years, supersymmetric extensions of integrable KP, KdV 
and NLS type hierarchies received much attention. This interest is motivated 
by both pure mathematical reasons and possible physical applications 
of these systems in non-perturbative $2D$ supergravity, matrix models, etc. 
One of the important motivations comes  
from the fact that these super-hierarchies are related, through 
their second hamiltonian structure, to superconformal algebras 
(both linear and $W$ type ones). Thus they provide additional insights 
into the theory of $W$ (super)algebras and conformal field theory. 
It is an urgent problem 
to fully classify all such systems and to reveal hidden 
relationships between them. 

Up to now, most efforts were focused on 
studying $N=1$ and $N=2$ supersymmetric systems (see, e.g., [1- 15]). 
Recently, an example of  
hierarchy with higher supersymmetry was found, the $N=4$ SKdV hierarchy 
\cite{{DI},{DIK}}. It admits the ``small'' $N=4$ SCA 
(with $SU(2)$ affine subalgebra) as the hamiltonian structure. 
In refs. \cite{{DG},{IK}} two different $N=2$ superfield scalar Lax 
formulations for this system were found. 

In this letter we demonstrate the existence of one more integrable 
SKdV hierarchy with the same $N=4$ SCA as the second hamiltonian
structure. Compared to the ``genuine'' $N=4$ SKdV, it possesses 
only $N=2$ rigid supersymmetry, and so does not possess 
a formulation in $N=4$ superspace. It should rather be viewed as an 
integrable 
extension of the $a=-2$, $N=2$ SKdV hierarchy by chiral and antichiral 
spin 1 $N=2$ superfields. One important difference with the $N=4$  
SKdV system is that, like the $a=-2$, $N=2$ SKdV, 
it admits a matrix Lax formulation 
(on the superalgebra $sl(3|2)$), in parallel 
with the scalar one. Since 
$N=4$ supersymmetry is broken to $N=2$ in this ``quasi'' 
$N=4$ SKdV system, its reductions to the systems having 
different $N=2$ subalgebras of the $N=4$ SCA as the second hamiltonian 
structures 
yield non-equivalent hierarchies. In this way we find two new 
integrable systems, both having the $N=2$ SCA as the second hamiltonian 
structure. 
One of them possesses only rigid $N=1$ supersymmetry and no $U(1)$ symmetry. 
The second possesses no supersymmetry, but respects $U(1)$ symmetry. 
It is still different from the non-supersymmetric system of ref. \cite{Mat1}. 
Thus the ``quasi'' $N=4$ SKdV system allows us to enlarge, 
via its reductions, the list of known fermonic extensions of KdV associated 
with the $N=2$ SCA. An unexpected peculiarity 
of the system constructed is that it is related, through a 
generalized Miura transformation, to one of the three $N=2$ super 
Boussinesq hierarchies \cite{{BIKP},{Yung2}}. 
This fact implies the existence of an intrinsic relationship between  
the ``small'' $N=4$ SCA and the nonlinear $N=2$ $W_3$ superalgebra.

\section{Matrix and scalar Lax representations}                   
We start by constructing the $N=2$ superfield matrix Lax operator 
which yields, through the appropriate Lax equation, the new SKdV system 
we intend to study in this paper. This will be done by applying 
the techniques of ref. \cite{DM} to the superalgebra $sl(3|2)$. We  
skip most details which will be given elsewhere. From now on, we 
deal with superfields on the $N=2$ superspace $X \equiv 
(x, \theta, \bar\theta)$ and use the following conventions 
\be
D=\frac{\partial}{\partial\theta} +
 \frac{1}{2}\bar\theta\frac{\partial}{\partial x},\quad 
\bar D=\frac{\partial}{\partial\bar\theta} +
 \frac{1}{2}\theta\frac{\partial}{\partial x},\;\;\;
\left\{ D,\bar D \right\} = \frac{\partial}{\partial x},\quad
D^2 = {\bar D}^2 =0\;.
\ee

In $N=2$ superspace, it is necessary to introduce two anticommuting 
Lax operators
\begin{equation}
{\cal L}=D+\Omega,\,\, \bar{\cal L}=\bar D+\bar\Omega,
\label{lax1}\end{equation}
where the connections $\Omega$ and $\bar\Omega$ take value in the 
loop algebra constructed from $sl(3|2)$. The loop parameter will be 
denoted by $\lambda$, and we choose the supertrace of 
a $5\times 5$ matrix $M$ to be
\begin{equation}
\mbox{str} M=M_{11}-M_{22}-M_{33}+M_{44}-M_{55}.
\end{equation}
Then one can take:
\begin{equation}
\Omega=\left(
\begin{array}{ccccc}
0 & 1 & 0 & 0 & 0 \\
0 & 0 & 0 & 0 & 0 \\
0 & 0 & 0 & 0 & 0 \\
DV & -V & \Phi_+ & 0 & 1 \\
0 & 0 & 0 & 0 & 0
\end{array}
\right) , 
\,\,
\bar\Omega=\left(
\begin{array}{ccccc}
0 & 0 & 0 & 0 & 0 \\
-V & 0 & 0 & 1 & 0 \\
\Phi_- & 0 & 0 & 0 & 0 \\
-\bar{D}V & 0 & 0 & 0 & 0 \\
\lambda & 0 & 0 & 0 & 0
\end{array}
\right),
\end{equation}
where $V$ is an unconstrained bosonic $N=2$ superfield, $\Phi_+$ and
$\Phi_-$ are, respectively, chiral, $D\Phi_+=0$, and antichiral,
$\bar D\Phi_-=0$, bosonic superfields. These connections satisfy 
the zero curvature equations
\begin{equation}
D\Omega+\hat\Omega\Omega=0,\,\,\bar D\bar\Omega+\hat{\bar\Omega}\bar\Omega=0,
\label{zero1}\end{equation}
where \^{} denotes the automorphism of the superalgebra which reverses 
the sign of odd generators. 

The Lax operators (\ref{lax1}) may be submitted to the commuting 
evolution equations
\begin{equation}
{\partial{\cal L}\over\partial t_k}= {\cal L}{\cal A}_k-
\hat{{\cal A}}_k{\cal L},\,\,\,
{\partial\bar{\cal L}\over\partial t_k}=\bar {\cal L}{\cal A}_k-
 \hat{{\cal A}}_k \bar{\cal L}  
\,\,\,\;.
\label{evol1}\end{equation}
The general construction of the matrices ${\cal A}_k$ follows the same lines
as in \cite{DM} and will not be given 
here. As an example which will be used in section 5, the explicit form 
of ${\cal A}_2$ 
$$ {\cal A}_2=-\left(
\begin{array}{ccccc}
\lambda+\Phi_+\Phi_- & 0 & -\bar D\Phi_+ & 0 & 0 \\
-\Phi_+D\Phi_- & \lambda+\Phi_+\Phi_- & \Phi_+' & 0 & 0 \\
D\Phi_-'-VD\Phi_- & -\Phi_-' & \Phi_+\Phi_- & -D\Phi_- & \Phi_- \\
\bar D\Phi_+ D\Phi_- & -\Phi_-\bar D\Phi_+ & -\bar D\Phi_+'-V\bar D\Phi_+
& \lambda+\Phi_+\Phi_- & 0 \\
0 & 0 & \lambda\Phi_+ & 0 & \lambda \end{array} \right).
$$
The equations (\ref{zero1}), (\ref{evol1}) are compatibility conditions 
for the linear problem
\begin{equation}
{\cal L}\Psi=0,\,\,\, \bar{\cal L}\Psi=0,\,\, {\partial\Psi\over
\partial t_k} +{\cal A}_k\Psi=0,
\label{linear}\end{equation}
where $\Psi$ is a $5$ component column vector. Through the usual 
elimination procedure, the first two of equations (\ref{linear}) 
lead to the following eigenvalue equation for $\Psi_5$
\begin{equation}
D\left( \partial_x +2\;V -\Phi_+ \partial^{-1} \Phi_- 
\right)\bar D \Psi_5=\lambda\Psi_5.
\end{equation}
Thus we arrive at the following $N=2$ scalar Lax operator and the Lax 
equation 
\be
L = D\left( \partial_x +2\;V -\Phi_+ \partial^{-1} \Phi_- \right) 
\bar D\;, \;\; \;\;\;\;\;\frac{\partial L}{\partial t_k} = -4 
\left[\; L^{{k\over 2}}_{\geq 0}, L \;\right]\;, 
\label{laxreprdif}
\ee
where the numerical coefficients and signs were chosen for 
further convenience.  

The explicit form of the third and second flow equations will be given 
below. Here we only note that in the limit $\Phi_{\pm} = 0$  
one recovers the $a=-2$, $N=2$ SKdV hierarchy in the formulation which uses a 
chirality-preserving Lax operator. Such scalar Lax 
representations first appeared in ref. \cite{Pop3} and were studied 
in more detail in \cite{DG}. 

Thus the new $N=2$ superhierarchy we have constructed is an 
integrable extension of the $a=-2$, $N=2$ SKdV heirarchy by chiral 
and anti-chiral $N=2$ superfields $\Phi_{\pm}$. In the next Section we will 
examine its hamiltonian structure. 
    
Note that the even conserved charges are given by the following 
expression
\be  \label{hscal}
H_k = \int \mu^{(2)}\; \mbox{Res}\; L^{{k\over 2}} 
\ee
where $\mu^{(2)} = dx d\theta d\bar \theta$ and the residue is 
defined as the coefficient before $D\bar D \partial^{-1}$. 

\setcounter{equation}{0}

\section{Hamiltonian formulation}
We could extract the second hamiltonian (or Poisson) structure associated 
with our system directly in the framework of the Lax 
representation, based on the formalism worked out in \cite{DG}. 
Here we do this in the hamiltonian framework. 

It will be instructive to consider this system in parallel with 
the $N=4$ SKdV system \cite{DI} which has the same $N=2$ 
superfield content \cite{DIK}. By introducing a parameter $c$ 
(not to be confused with the central charge of the $N=2$ and $N=4$ SCA), 
we can uniformly write the third flow equations of both hierarchies as   
\bea
\frac{\partial V}{\partial t_3} &=& - V ''' + 3\; 
\left( \left[ D, \bar{D} \right] V 
\;V \right) '
+ {1\over 2}(5-2c)\;\left( \left[ D,\bar{D} \right] V^2 \right) ' 
\nonumber \\
&&+ 2(c-3)\; \left( V^3 \right) '  
+ (c-1) \; \left( \Phi_- \Phi_+ ' - 
\Phi_+ \Phi_- '
\right) ' \nonumber \\ 
&& +  6 \left( V\Phi_+ \Phi_- \right) ' +
{1\over 2}(c-4)\left( D\Phi_- \bar D \Phi_+ \right) '\;, \label{Veqc} \\
\frac{\partial \Phi_+}{\partial t_3} &=& -c \;\Phi_+ '''- 6\; 
D \left[  \bar D \Phi_+ V ' 
 + {1\over 3}\;(c+2)\;\bar D \Phi_+ ' V \right. \nonumber \\ 
&& + \left. {4\over 3}\;(4-c)\;
V\bar DV \Phi_+ 
- \bar D \Phi_+ \left( \Phi_+\Phi_- + {1\over 3}\;(c-7)\;V^2
\right) \right]\;. \label{phi2c}
\eea 
(the equation for $\Phi_-$ can be restored through the discrete 
automorphism $\Phi_{\pm} \leftrightarrow \Phi_{\mp}, V \rightarrow -V, 
D \leftrightarrow \bar D$ which is a symmetry of both hierarchies).
At $c=4$ we get just the system constructed here; at $c=1$ the 
manifestly $U(1)$ symmetric $a=4, b=0$ ``gauge'' of 
$N=4$ SKdV in the $N=2$ superfield form \cite{DIK} is recovered. 
This notation emphasizes not only 
the similarity but also an essential difference between both systems: 
in the limit $\Phi_{\pm} =0$, the second one goes into the $a=4$, $N=2$ 
SKdV system
\footnote{Actually, the $a=-2$, $N=2$ SKdV hierarchy can also be obtained as 
a consistent reduction of the $N=4$ SKdV one, but with another choice of  
``gauge'' with respect to the broken $SU(2)$ automorphism symmetry of $N=4$ 
supersymmetry \cite{DIK}. This property will be recovered in another 
context below.}. 

Let us write the conserved dimension 3 
hamiltonian 
for the system \p{Veqc} - \p{phi2c}
\be 
H^c_3 =  \int \mu^{(2)} \left\{
[D, \bar D]V \;V  + {2\over 3} \;(c -3) V^3 + 2\;V \Phi_+\Phi_-  
 + {c\over 2}\;\Phi_+ '\Phi_- \right\} \label{hamc}
\ee
(for our case we made use of the general formula \p{hscal}, 
while in the $N=4$ SKdV case it was given 
in ref. \cite{DIK}). Now it is a matter of straightforward computation 
to show that the set \p{Veqc} - \p{phi2c} 
can be given the hamiltonian form 
\be
\frac{\partial V^A}{\partial t_3} = \{ V^A, H^c_3 \} = 
{\cal D}^{AB} \frac{\delta H^c_3}{\delta V^{B}}\;, 
\;\;\;
V^{A} \equiv \left( V, \Phi_-, \Phi_+ \right)\;, 
\label{hamform}
\ee
with the following Poisson brackets algebra
\bea
&& \left\{ V^A(1), V^B(2) \right\} =  {\cal D}^{AB}(1) 
\Delta^{(2)}(1-2)\;,
\label{poi2} \\
&& {\cal D}^{11} = \partial V - (\bar D V) \;D 
- (DV) \; \bar D - {1\over 2}\; [D,\bar D] \partial \;,  
\nonumber \\
&& {\cal D}^{21} = 
 \bar D \left( D\Phi_- + \Phi_- D \right)\;,
\;\;\;
{\cal D}^{31} = 
 D \left( \bar D \Phi_+ + \Phi_+ \bar D \right)\;, \nonumber \\
&& {\cal D}^{23} = 
2 \bar D\left( \partial - 2V \right) D \;,\;\;\; 
{\cal D}^{22} = {\cal D}^{33} = 0 \;.\label{poi2det}
\eea
Here $\Delta^{(2)} (1 - 2) = 
\delta(x_1 -x_2) 
(\bar \theta_1 - \bar \theta_2)(\theta_1 - \theta_2)$. The remaining 
entries of the operator ${\cal D}^{AB}(1)$ can be easily deduced  
from those in \p{poi2det}.

The above Poisson superalgebra is the classical ``small'' 
$N=4$ SCA (at some fixed value of the central 
charge) already used to construct the $N=4$ SKdV hierarchy 
\cite{{DI},{DIK}}. Thus we have explicitly shown that the new hierarchy 
is also associated with this superalgebra as the second hamiltonian structure. 
The lack of rigid $N=4$ supersymmetry in our $c=4$ case and its presence 
in the $N=4$ SKdV ($c=1$) case can be readily established by 
considering the transformation properties of eqs. \p{Veqc} - \p{phi2c} 
and of the hamiltonian 
\p{hamc} under the hidden $N=2$ supersymmetry \cite{DIK}
\bea
\delta V = {1\over 2}\;\hat{\epsilon} \;D\Phi_- + 
{1\over 2}\;\hat{\bar{\epsilon}} \;
\bar{D} \Phi_+ \;, \;\;
\delta \Phi_- = - 2\;\bar{\hat{\epsilon}} \bar{D} V\;, \;\;
\delta \Phi_+ =  -2\;\hat{\epsilon}\; D V \;,
\label{susy2} \eea
which extends the manifest $N=2$ supersymmetry to $N=4$.    
Under these variations the 
integrand in \p{hamc} is shifted by full spinor derivatives at $c=1$. 
This is not the 
case for any other choice of $c$. The absence of $N=4$ supersymmetry 
in the set \p{Veqc} - \p{phi2c} 
at $c=4$ is obvious already from the fact that the linear terms in the 
r.h.s. of these equations appear with inequal coefficients. One more way 
to see the same property is to note that the generators of the 
hidden $N=2$ supersymmetry 
$$
Q \sim \int \mu^{(2)}\bar\theta \Phi_-\;,\;\;\bar Q \sim \int 
\mu^{(2)} \theta \Phi_+
$$
are conserved at $c=1$, but not at $c=4$ . To 
summarize, our system can be viewed as an integrable extension of 
the $a=-2$, $N=2$ super KdV hierarchy, such that it preserves the $N=2$ 
supersymmetry and $U(1)$ symmetry of the latter and possesses the "small" 
$N=4$ SCA as the second hamiltonian structure. 

One of the basic attributes of integrability is the existence of 
an infinite set of mutually commuting conserved charges. For the $N=4$ 
SKdV this property has been first demonstrated in \cite{DI}, \cite{DIK} 
by the explicit computation of the conserved bosonic charges up to the 
dimension 6 and proving that this system is bi-hamiltonian. Now, after 
constructing Lax operators for this system in refs. \cite{{IK},{DG}}, the 
existence of an infinite set of conserved quantities for it is a consequence 
of the appropriate Lax representation. The same is true of our 
"quasi" $N=4$ super KdV system: 
the bosonic conserved quantities of any dimension can be computed according 
to the general formula \p{hscal} (or by a similar one in the matrix 
Lax formulation). To illustrate the general procedure, 
we quote the expressions for the 
conserved charges of dimension 2 and 4. For the purpose of comparing 
with the $N=4$ SKdV case, we again write them in parallel for both systems 
\bea
H_{2}^c &=& \int \mu^{(2)} \left\{ \Phi_+\Phi_- + 
{2\over 3}(c-4)V^2 \right\}\;, 
\label{h2c} \\
H_{4}^c &=& \int \mu^{(2)} 
\left\{ VD\Phi_-\bar D\Phi_+ + {1\over 4} \Phi_-^2 \Phi_+^2 
- {1\over 2}\Phi_- \Phi_+ '' 
+ {2\over 9}(4-c) \left[ V^4 +{1\over 2}VV '' \right. 
\right. \nonumber \\
&& - \left. \left. {3\over 2} V^2[D,\bar D]V 
- 3V^2\Phi_+\Phi_- 
+{1\over 2}\Phi_-\Phi_+ '' \right]  \right\}\;. \label{h4c}
\eea
We see that, as opposed to the $N=4$ SKdV case, the integrands in 
these charges at $c=4$ are vanishing in the $N=2$ SKdV 
limit $\Phi_{\pm} = 0$. Actually, this property extends to all the 
charges of even dimensions. It matches with the well-known fact that the 
$a=4$, $N=2$ SKdV possesses higher-order bosonic conserved quantities 
of all integer dimensions, while for the $a=-2$ system only 
odd dimension charges exist \cite{{Mat1},{Mat2}}. 

Note that the $N=4$ SKdV system, like the $a=4$, $N=2$ one,  
possesses a first hamiltonian structure which is local and linear, 
the charge $H_4^{c=1}$ being the relevant hamiltonian \cite{{DI},{DIK}}. 
It is obvious from the form of $H_4^{c=4}$ that no such first hamiltonian 
structure can be defined for the ``quasi'' $N=4$ SKdV system, at least 
in the class of polynomial and local hamiltonians ($H_4^{c=4}$ contains no 
terms bilinear in $V$, so there is no way to reproduce the linear term 
in the r.h.s of eq. \p{Veqc} at $c=4$).  
  
\setcounter{equation}{0}

\section{Integrable reductions}  
The interplay between the above $c=1$ and $c=4$ SKdV 
systems resembles the well-known relation between the $N=1$  
SKdV hierarchy \cite{{MR},{a4},{Mat}} and the non-supersymmetric 
integrable fermionic extension of KdV constructed in \cite{{Kup},{a4}}. 
Despite the radically 
different symmetry properties, 
both these systems have $N=1$ SCA as 
the second hamiltonian structure. They also admit a 
unifying parametrization by the parameter $c$, with $c=1$ for the 
$N=1$ supersymmetric system and $c=4$ 
for the non-supersymmetric one. Analogously, the $N=2$ SCA 
gives rise to two essentially 
different kinds of integrable extensions of KdV: three $N=2$ supersymmetric 
ones and a non-supersymmetric one \cite{Mat1}. In our case we 
encounter a similar situation: two different integrable 
systems prove to be associated with the same $N=4$ SCA
as the second hamiltonian structure. One possesses full global $N=4$ 
supersymmetry, 
while another has only $N=2$ supersymmetry. We conjecture that this is 
a general phenomenon. Namely, for each of these superconformal 
algebras one can define a whole sequence of integrable fermionic 
extensions of KdV, ranging from the systems with the maximally 
possible global supersymmetry to the non-supersymmetric 
systems. If this conjecture is true, 
then, beginning from $N=2$, the intermediate integrable systems 
should exist, 
in which the maximal supersymmetry is broken only partially. The 
existence of the above ``quasi'' $N=4$ SKdV hierarchy 
with $N=2$ supersymmetry confirms this hypothesis. One may wonder 
why in the $N=2$ case only two kinds of integrable 
extensions of KdV are known. 
Now we wish to show that in fact two more integrable fermionic extensions 
of KdV equation with $N=2$ superconformal algebra as the second 
hamiltonian structure exist. Both are self-consistent reductions of 
our ``quasi'' $N=4$ super KdV system. The first of them possesses 
only $N=1$ supersymmetry and no internal $U(1)$ symmetry. 
The second yields a non-supersymmetric, though $U(1)$ 
symmetric, system which is still different from the extension constructed 
in ref. \cite{Mat1}.

For our aim it will be convenient to deal with the equivalent 
$N=1$ superfield form of the ``quasi'' $N=4$ super KdV system. Passing to 
new $\theta$'s , $\theta^1 = {1\over 2}(\theta +\bar \theta)$, 
$\theta^2 = {1\over 2}(\theta -\bar \theta)$ (these are, respectively, 
real and imaginary with respect to the involution 
$\theta \leftrightarrow \bar \theta $), and 
redefining appropriately the spinor derivatives 
\bea
D = {1\over 2} (D_1 +D_2)\;, \; \bar{D} = {1\over 2} (D_1 -D_2), \;
(D_1)^2 = - (D_2)^2 = \partial, \;
\{ D_1, D_2 \} = 0\;,  
\label{D1D2}
\eea
we can express the $N=2$ superfields $V (X), \Phi_\pm (X)$ through 
$N=1$ ones depending on $(x,\theta^1)$ 
\be 
V = J_0 + \theta^2 \;G\;, \; \Phi_{\pm} = J_{\pm} \mp 
\theta^2 \;D_1 J_{\pm}\;.  \label{defN2kdv} 
\ee
The system \p{Veqc} - \p{phi2c} at $c=4$ can be rewritten in 
$N=1$ superspace as follows 
(from here on, we omit the index 1 on $N=1$ spinor derivative and $\theta$) 
\begin{eqnarray}
\frac{\partial G}{\partial t_3} &=& - G '''- 3\left(GDG \right)'+ 6 
\left[ G \left(J_+J_-+J_0^2 \right) + J_0 
\left( J_+DJ_- - J_-DJ_+ \right) \right]'
\nonumber\\
&&+ \;3 \;\left(J_{0}' DJ_0 
-J_+DJ_- ' -J_-DJ_+ ' + J_+ ' DJ_- + J_-' DJ_+ \right)' \;,
\\ \frac{\partial J_0}{\partial t_3} &=& - J_0 ''' + 2 
\left(J_0^3\right)' 
+ 3 \left(J_-J_+' - J_+J_- '- G\; DJ_0 \right) ' + 6 
\left(J_0J_+J_-\right) ' \;,\\ 
\frac{\partial J_+}{\partial t_3} &=& - 4\;J_+ '''- 12\; J_0J_+ '' - 
6 \;J_0 ' J_+ '
- 6 \;J_0^2J_+ ' + 6 \;DJ_+ ' DJ_0 -3\;DJ_0 ' DJ_+ \nonumber\\
&& + 6 \left(J_+J_-J_+ ' +J_+DJ_-DJ_+ -J_0DJ_0DJ_+ \right)
+ 6 \left(DJ_+ ' +J_0 DJ_+\right)G  \nn \\
&&- 3 G '\;DJ_+  \label{N1sf}
\end{eqnarray}
(the equation for $J_-$ can be obtained from that for $J_+$ through 
the appropriate involution). The $N=1$ superspace form of the Lax 
representation is as follows 
\be  \label{N1Lax}
L = \partial^2 + J_0 \partial +DJ_0 D + GD -J_+\partial^{-1}DJ_-D\;,
\;\;\; \;\;\;\;
\frac{\partial L}{\partial t_k} = -4 
\left[\; L^{k\over 2}_{>0}, L \;\right]\;.
\ee

The hamiltonian \p{hamc} in terms of $N=1$ superfields reads
\bea 
H^c_3 &=& {1\over 2} \int dxd\theta \left\{ GDG - DJ_0 J_0 '  - 
{c\over 2} \left[ 
(J_+ ' DJ_- + J_- ' DJ_+ \right] \right. \nonumber \\ 
&& + \left. 2G \left[ (3-c)\;(J_0)^2 - J_+J_- \right] + 2J_0 
\left[ J_- DJ_+ 
- J_+ DJ_- \right] \right\}\;.\label{hamcN1}                 
\eea   
The standard reduction 
$\Phi_{\pm} = 0$ amounts to putting 
\be
J_{\pm} = 0 \label{stanred} 
\ee
in \p{hamcN1}. This yields 
the $a=4$ and $a=-2$, $N=2$ super KdV hierarchies for $c=1$ and $c=4$, 
respectively, with the same second hamiltonian structure 
$N=2$ SCA generated by 
the $N=1$ superfields $G$ and $J_0$. However, one may 
embed the $N=2$ SCA into the $N=4$ SCA in different ways, and perform the 
reduction of our system so as to finally have another $N=2$ subalgebra 
of $N=4$ SCA as the hamiltonian 
structure. For the $c=1$ case all such reductions yield $N=2$ 
supersymmetric systems because the initial system is 
$N=4$ supersymmetric. On the other hand, 
for the $c=4$ case this is no longer true because the 
$c=4$ system possesses 
only $N=2$ rigid supersymmetry. It is just the one with respect to 
which $G$ and $J_0$ form an irreducible multiplet and which 
is manifest in the formulation exposed in 
the previous sections. The $c=4$ system is not invariant 
with respect to any other $N=2$ subsymmetry of the whole 
rigid $N=4$ supersymmetry formed by 
the manifest $N=2$ supersymmetry transformatons and 
those given by \p{susy2}. 

If we wish to preserve the $N=1$ superfield structure and hence $N=1$ 
supersymmetry in the process of 
reduction, then, beside 
the $N=2$ supersymmetry which is manifest in the previous $N=2$ 
superfield formulation, only two other appropriate $N=2$ supersymmetry 
subalgebras exist. They are formed by the explicit 
$N=1$ supersymmetry transformations combined with 
the ``real'' or 
``imaginary'' parts of the hidden $N=2$ supersymmetry 
transformations \p{susy2}. These parts correspond to singling out the  
following combinations of the parameters $\hat{\epsilon}$, 
$\bar{\hat{\epsilon}}$ 
$$
\hat{\epsilon}_1 = {1\over 2}(\hat{\epsilon} + \bar{\hat{\epsilon}})\;, \;\;
\hat{\epsilon}_2 = {1\over 2}(\hat{\epsilon} - \bar{\hat{\epsilon}})\;.
$$
With respect to these two different $N=2$ 
supersymmetries the $N=1$ superfields 
$$
G,\; J_0, \;J_- \equiv J_1 + J_2, \; J_+ \equiv J_1 - J_2\;, 
$$
fall into the following two sets of $N=2$ supermultiplets 
\be
(1). \;\;\left( G, \; J_1 \right)\;, \;\; \left( J_0, \; J_2 \right)\;;   
\;\;\;\;\; (2). \;\; \left( G, \; J_2 \right)\;, \;\; 
\left( J_0, \; J_1 \right) \;.
\label{N2mult}
\ee
It is easy to check that each of these pairs is indeed closed under 
the appropriate $N=2$ supersymmetry, e.g., 
\be
\delta_{\hat{\epsilon}_2}G  = \hat{\epsilon}_2 J_1'\;, \; 
\delta_{\hat{\epsilon}_2} J_1 = - \hat{\epsilon}_2 G\;, \; 
\delta_{\hat{\epsilon}_2} J_0 = \hat{\epsilon}_2 DJ_2\;, \; 
\delta_{\hat{\epsilon}_2} J_2 =  \hat{\epsilon}_2 DJ_0\;.
\label{hidtr}
\ee 
Let us now perform the reduction 
\be 
J_0 = J_2 = 0\;,  \label{newred}
\ee  
which brings \p{hamcN1} into 
\be
H^c_{red} = \int dxd\theta \left[\; GDG - c\;DJ_1 J_1' - 
2\;G (J_1)^2 \;\right] \;.
\label{hamred}
\ee
Note that, like the reduction \p{stanred}, this reduction is 
self-consistent in the sense that both left- and right-hand sides of 
the equations for $J_0$ and $J_2$ disappear in 
this limit (for any flow with odd scaling dimension). Therefore, 
the resulting systems inherit the integrability 
properties of the initial systems, in particular, the presence of 
infinite sets of conserved charges. It can be easily checked that the 
Poisson brackets of the remaining superfields $G, J_1$ form the $N=2$ SCA 
which is isomorphic to but different from the $N=2$ SCA generated 
by $G$ and $J_0$. 

Looking at \p{hamred} we observe that at $c=1$ this is 
none other than the hamiltonian of the $a=-2$, $N=2$ SKdV 
hierarchy written in terms of $N=1$ superfields. It is easy to check 
its invariance both under $N=2$ supersymmetry  \p{hidtr} and 
$U(1)$ transformations. The latter are realized on $G$ and $J_1$ as 
\be 
\delta J_1 = \alpha \theta^1 G\;, \; \delta G = \alpha 
\frac{\partial}{\partial \theta^1} J_1\;, \label{u1}
\ee
$\alpha$ being a constant parameter.
The fact that the reductions 
\p{stanred} and \p{newred} of the same $a=4, b=0$ $N=4$ super KdV equation 
yield the $a=4$ and $a=-2$, $N=2$ SKdV equations demonstrates, in 
a slightly different fashion than in \cite{DIK}, that both these 
inequivalent $N=2$ SKdV hierarchies are contained 
as particular solutions in the single $N=4$ SKdV one. 
The $N=2$ structures $G, J_0$ and $G,J_1$ and, respectively, 
the reductions \p{stanred} and \p{newred} are related to each other by 
some global $SU(2)$ transformation. The fact that they give rise to 
different $N=2$ KdV systems comes from the non-invariance of the $N=4$ 
SKdV hamiltonian with respect to the full set of such $SU(2)$ 
rotations \cite{DI}, \cite{DIK}.       

At $c=4$ the hamiltonian \p{hamcN1} respects neither $N=2$ supersymmetry 
nor $U(1)$ invariance \p{u1}  
\be
H^{c=4}_{red} = \int dxd\theta \left[ \;GDG - 4 \;DJ_1 J_1' - 2\;G (J_1)^2
\;\right]\;. \label{N1c4}
\ee
Its only residual symmetry is $N=1$ supersymmetry. Thus we have derived 
a new integrable extension of KdV equation which still has $N=2$ SCA as 
the second hamiltonian structure, enjoys $N=1$ supersymmetry but 
possesses no $U(1)$ invariance. It is instructive to present the relevant 
equations which follow from \p{N1sf} upon imposing \p{newred}
\bea 
\frac{\partial G}{\partial t_3} &=& -G''' -3 \;\left( DG G \right)' + 
6\;\left( G J_1^2  
+ J_1'DJ_1 - J_1 DJ_1' \right)'  \label{Qeqred} \\
\frac{\partial J_1}{\partial t_3} &=& -4\; J_1''' + 
2\left( J_1^3 \right)' + 6\;DJ_1' G + 3 DJ_1 G' \;. \label{Jeqred}
\eea
The appropriate $N=1$ superfield Lax operator and Lax equations 
can be obtained by substituting \p{newred} into \p{N1Lax} and 
restricting $k$ in \p{N1Lax} to odd integers, $k=2n+1$.  

The absence of the second supersymmetry stems from 
the fact that its generator $\int dxd\theta J_1$ is not conserved. 
An interesting property of this reduction is that the 
even dimension conserved charges $H_2$ and $H_4$ \p{h2c} and \p{h4c} 
disappear, like in the $c=1$ case. This property 
extends to the whole sequence of the even dimension charges and is related to 
the invariance of our ``quasi'' $N=4$ SKdV hierarchy 
under the discrete automorphism 
\begin{equation}
J_0\rightarrow -J_0,\,J_+\rightarrow J_-,\,J_-\rightarrow J_+,\, 
G\rightarrow G\;.
\end{equation}
It preserves the odd-dimension conserved charges but reverses 
the sign of the even-dimension ones. This property and the fact that 
the reduction \p{newred} is a fixed point of this discrete symmetry, 
explain the vanishing 
of the even-dimension charges in the case at hand (like 
in the case of $a=-2$, 
$N=2$ SKdV).
     
We can choose the $N=2$ SCA in the underlying $N=4$ SCA 
so that the reduction associated with this choice yields a system 
having {\it no supersymmetry at all}. Let us go to the component 
fields 
\be 
G= \xi_1 + \theta^1 u \;, \;J_0 = j_0 + \theta^1 \xi_2\;, \; 
J_1 = j_1 + \theta^1 \xi_3\;, \; J_2 = j_2 + \theta^1 \xi_4 \;,
\label{comp} 
\ee
where all fields are functions of $t$ and $x$. We wish to examine the 
multiplet structure of these fields with respect to the $N=2$ 
supersymmetry \p{susy2} which is not a symmetry of the 
hamiltonian \p{hamcN1} at $c=4$ (though it is at $c=1$). With respect to the
transformations with parameters $\hat{\epsilon_2}$ and $\hat{\epsilon_1}$ 
these fields are split, respectively, into the following 
irreducible $N=1$ multiplets 
\bea 
&& N=1\; (\hat{\epsilon_2}): \;\;\;\; \left( \xi_3, u \right)\;, \;\; 
\left(j_0,  \xi_4 \right)\;, \;\; \left( j_1,  \xi_1 \right)\;, \;\; 
\left( j_2, \xi_2 \right) \label{N11} \\
&& N=1\; (\hat{\epsilon_1}): \;\;\;\; \left( \xi_4, u \right)\;, \;\; 
\left(j_0,  \xi_3 \right)\;, \;\; \left( j_2,  \xi_1 \right)\;, \;\; 
\left( j_1, \xi_2 \right)\;.  \label{N12}
\eea
The first two and last two pairs in each sequence form 
$N=2$ multiplets. One can check that the $N=2$ multiplet  
\be 
 \left( j_0, \xi_3, \xi_4, u \right) 
\ee
generates an $N=2$ SCA. Then one can enforce the reduction which yields an  
extension of KdV system with this particular $N=2$ SCA as the second 
hamiltonian structure  
\be
j_1 = j_2 = \xi_1 = \xi_2 = 0\;. \label{red3}
\ee 
It is a simple exercise to be convinced that this 
reduction is also consistent: evolution equations for the fields in 
\p{red3} are identically satisfied when we impose \p{red3}. 

When $c=1$, due to $N=4$ supersymmetry of the hamiltonian, one ends up 
again with an $N=2$ supersymmetric integrable system, namely, 
the $a=-2$, $N=2$ 
SKdV. A radically different situation comes out when $c=4$. It is easy to 
see that in this case \p{red3} explicitly breaks 
the whole supersymmetry of hamiltonian \p{hamcN1}, since
these constraints are covariant 
under $\hat{\epsilon}_{1,2}$ supersymmetries only, which are not 
respected by this hamiltonian. At the same time, they are 
covariant under the $U(1)$ symmetry which mixes $j_1$ with $j_2$ and 
$\xi_1$ with $\xi_2$. As a result, the reduced system should also be 
$U(1)$ covariant. The reduced hamiltonian is as follows 
\be
H^{c=4}_{red '} = {1\over 2}\int dx \left\{  u^2 + 
4 \left(\xi_3 - \xi_4 \right) \left(\xi_3 + \xi_4 \right)'  - j_0' j_0'  
- 2 u (j_0)^2 + 8 j_0 \xi_3 \xi_4 \right\}.\label{hamred3}
\ee 
Its $U(1)$ (or $SO(1,1)$, depending on which reality properties are 
ascribed to the fields) symmetry realized by proper rescalings of 
the fermionic fields  is manifest. 

Thus we have got one more new integrable model with $N=2$ SCA as 
the second hamiltonian structure algebra. It possesses no supersymmetry but 
respects global $U(1)$ invariance. It differs from the 
 bi-hamiltonian non-supersymmetric $U(1)$ invariant 
fermionic extension of KdV with the
$N=2$ SCA second hamiltonian structure found in \cite{Mat1}. 
The main difference between 
both systems is the equation for the $U(1)$ current. In the system 
of ref. \cite{Mat1} it satisfies the trivial equation 
$\frac{\partial j_0}{\partial t_3}= 0$, while in our case it satisfies 
the mKdV equation 
$$
\frac{\partial j_0}{\partial t_3} = - j_0''' + 2\;(j_0^3)'\;.
$$ 
Thus the $j_0$ equation decouples in both systems, but in different ways. 
The change of variables \cite{Mat1}
$q = u -(j_0)^2$, $\psi_\pm = \mbox{exp}\{\pm \partial^{-1}j_0 \} 
(\xi_3 \mp \xi_4)$ fully separates the $j_0$ and $q, \psi_{\pm}$ equations, 
the latter set proving to be the same as in ref. \cite{Mat1}. 
The analysis in \cite{Mat1} was essentially bound by requiring the existence 
of a bi-hamiltonian structure, while our system certainly possesses 
no local first 
hamiltonian structure. The $N=1$ superfield Lax formulation \p{N1Lax} 
under the reduction \p{red3} gives rise to the two independent 
component Lax operators
\be \label{red2lax}
L^{(1)} = \partial^2 + 2j_0\partial\;,\;\; \;\;\;\;\;L^{(2)} 
= \partial^{2} + q - \psi_+\partial^{-1}\psi_-\;,
\ee
each producing its own hierarchy. After putting 
$j_0 = \xi_4 =0$ this system, like the one constructed in \cite{Mat1}, goes 
into the non-supersymmetric fermionic extension of 
KdV with the $N=1$ SCA as the second hamiltonian structure \cite{{Kup},{a4}}.   
\setcounter{equation}{0}

\section{Relation to the $N=2$ Boussinesq hierarchy}
As the last topic, we quote a surprising relationship between  
our ``quasi'' $N=4$ SKdV system and the $\alpha = -2$, $N=2$ super 
Boussinesq hierarchy.  

We will establish this relationship at the level of the second flows. 
For the ``quasi'' $N=4$ SKdV the corresponding equations can be 
straightforwardly derived, e.g.,  through
the Poisson structure \p{poi2}, \p{poi2det} with 
$H_{2}^{c=4}$ \p{h2c} as the hamiltonian 
\be
\frac{\partial V_A}{\partial t_2} = -{1\over 2} \left\{V_A, H_2 \right\}
\ee
(the numerical coefficient was chosen for further 
convenience). Their explicit form is 
\bea 
\frac{\partial V}{\partial t_2} &=& -(\Phi_+\Phi_-)' \;, \nn \\
\frac{\partial \Phi_+}{\partial t_2} &=& 2 \left( \Phi_+'V +
DV \bar D \Phi_+  + 
{1\over 2} \Phi_+'' \right), \; 
\frac{\partial \Phi_-}{\partial t_2} = 2 \left( \Phi_-' V + 
\bar D V D \Phi_-  - 
{1\over 2} \Phi_-'' \right). \label{2n4q}
\eea 

Now, let us assume that at least one of the superfields $\Phi_{\pm}$ 
is invertible (i.e. starts with a constant) and define the following 
Miura type transformations 
\bea 
\tilde V_1 = V  + \frac{\Phi_+'}{\Phi_+}\;, \;\; 
W_1 = \Phi_+\Phi_- - 2 \left( \frac{\Phi_+'}{\Phi_+} V + 
DV \frac{\bar D \Phi_+}{\Phi_+}  +{1\over 2}\frac{\Phi_+''}{\Phi_+} \right)
\label{M1}  
\eea 
or 
\bea 
\tilde V_2 = V  - \frac{\Phi_-'}{\Phi_-}\;, \;\; 
W_2 = \Phi_+\Phi_- + 2 \left( \frac{\Phi_-'}{\Phi_-} V + 
\bar D V\frac{D \Phi_-}{\Phi_-}  -{1\over 2}\frac{\Phi_-''}{\Phi_-} \right)\;.
\label{M2} 
\eea
It is easy to show that the spin 1 
superfield $\tilde V$ and the composite spin 2 $W$ satisfy, as 
a consequence of eqs. \p{2n4q}, the following set of equations 
\bea
\frac{\partial \tilde{V}}{\partial t_2} = -W'\;,\;\;
\frac{\partial W}{\partial t_2} = -[D,\bar D] W' +2 \left(\tilde{V} 
W' +D\tilde{V} \bar D W  + \bar D\tilde{V} D W  \right)\;.
\eea
This system is recognized as the second flow of 
the $\alpha = -2$ $N=2$ Boussinesq hierarchy \cite{{BIKP},{Yung2}} (in the 
classification of ref. \cite{BIKP}). The same relation can be established 
for any flow and for the relevant Lax operators. Instead of 
presenting it explicitly here (it is a particular case of 
a general relationship between different families of Lax operators 
in $N=2$ superspace \cite{DG}), we will illustrate it on the examples 
of the conserved charges $H_2$, $H_3$ and $H_4$. Namely, the Miura 
transformations \p{M1} or \p{M2} map the expressions \p{h2c}, \p{hamc} 
and \p{h4c} at $c=4$ on the following ones (up to rescalings)
\bea
H_2^b = \int \mu^{(2)} W,\;  
H_3^b  =  \int \mu^{(2)} \left( [D,\bar D] \tilde{V}\;\tilde{V} 
+ {2\over 3} \tilde{V}^3 + 2 \tilde V W \right),\; 
H_4^b = \int \mu^{(2)} W^2\;. \label{hbou}
\eea   
These are just the conserved charges of the $\alpha = -2$, $N=2$ Boussinesq 
hierarchy. 

Thus we have explicitly constructed the generalized Miura map relating 
our ``quasi'' $N=4$ SKdV hierarchy to one of $N=2$ Boussinesq 
hierarchies. Such a map implies the existence of an intrinsic relationship 
between the second hamiltonian structures of both 
hierarchies: the ``small'' $N=4$ SCA and the $N=2$ $W_3$ algebra. 
Note that 
similar Miura type transformations were considered in 
refs. \cite{{KS},{KST},{BKS}}. 

This relationship provides a link between 
the $\alpha = -2$, $N=2$ Boussinesq and $a=-2$, $N=2$ SKdV hierarchies.  
A consistent reduction of the ``quasi'' $N=4$ SKdV hierarchy is
\be
\Phi_- = 0, \;\;\Phi_+ = const\;,\;\;\; \mbox{or} \;\;\; \Phi_- = const, 
\;\; \Phi_+ = 0\;.
\ee
All the even-dimension conserved charges vanish while 
the odd-dimension ones go into those of the $a=-2$, $N=2$ SKdV hierarchy. 
The same conditions imply the vanishing of one of the composite spin 
2 superfields $W_{1,2}$ introduced by eqs. \p{M1}, \p{M2}. So from  
the $N=2$ Boussinesq viewpoint such a reduction amounts to putting 
equal to zero the superfield $W$. Thus $W=0$ is a consistent reduction 
of $\alpha = -2$, $N=2$ Boussinesq hierarchy and takes it into 
the $a=-2$, $N=2$ SKdV one.  
    
\setcounter{equation}{0}
\section{Conclusions}
In this paper we have presented one more $N=2$ SKdV 
hierarchy with the ``small'' $N=4$ SCA as the second hamiltonian structure. 
We constructed for it both matrix and scalar Lax formulations. 
As compared to the ``genuine'' $N=4$ SKdV hierarchy, the new one respects 
only $N=2$ supersymmetry. Thanks to the partial breaking of $N=4$ 
supersymmetry, there are possible non-trivial consistent reductions 
of this system which yield previously unknown SKdV hierarchies with 
lower supersymmetry. In this way we found two new fermionic extensions 
of the KdV hierarchy, both having the $N=2$ SCA as the second 
hamiltonian structure.
One of them possesses $N=1$ supersymmetry, the second one is a new 
non-supersymmetric extension. The existence of 
such ``horizontal'' hierarchies of integrable 
SKdV systems having the same hamiltonian structure and ranging from 
maximally supersymmetric systems to the systems with completely 
broken supersymmetry seems to be a general phenomenon. If this conjecture 
is true, then there should exist even more SKdV hierarchies associated 
with $N=4$ SCA, $N=1$ supersymmetric and non-supersymmetric ones. 
It would be interesting to check this. 

One more notable property of the system presented here is its intimate 
relationship with the 
$\alpha = -2$, $N=2$ super Boussinesq one. We explicitly constructed 
a generalized Miura transformation mapping the first hierarchy on the 
second one. This map also relates the two relevant second hamiltonian 
structures, the ``small'' $N=4$ SCA and the $N=2$ $W_3$ superalgebra, thus 
revealing hidden links between these two superconformal algebras. 
One may think about possible implications of this remarkable relationship 
in $N=2$ $W_3$ strings, say.  

\vspace{0.5cm}

\noindent{\Large\bf Acknowledgement} 
\vspace{0.3cm}

\noindent E.I. is grateful to S. Krivonos for useful discussions. 
He also thanks ENSLAPP, ENS-Lyon, for the hospitality extended to him 
during the course of this work. His work was supported by the grant 
of Russian Foundation of Basic Research RFBR 96-02-17634, by INTAS 
grant INTAS-94-2317 and by a grant of the Dutch NWO organization. The work
of F.D. was supported in part by a grant from the HCM program of the European
Union (ERBCHRXCT920069).

\end{document}